\documentclass[preprint,secnumarabic,amssymb, nobibnotes, aps, prb,superscriptaddress]{revtex4-1}
\usepackage{amssymb,amsmath}
\usepackage{amssymb}

\newcommand{\one}{{(1)}}
\newcommand{\two}{{(2)}}
\newcommand{\makevec}[1]{{\bf #1}}
\newcommand{\uvec}{\makevec{u}}
\newcommand{\qvec}{\makevec{q}}
\newcommand{\Rvec}{\makevec{R}}

\newcommand{\vvec}{\makevec{v}}
\newcommand{\chivec}{\makevec{\boldsymbol \chi}}
\newcommand{\epsvec}{\makevec{\boldsymbol \epsilon}}
\newcommand{\xivec}{\makevec{\boldsymbol \xi}}
\usepackage{graphicx}

\begin{document}

\title{R-matrix theory for nanoscale phonon thermal transport across devices and interfaces}
\author{K.G.S.H. Gunawardana}
\email[Electronic address:]{harsha@ou.edu}
\author{Kieran Mullen}
\affiliation{Homer L. Dodge Department of Physics and Astronomy, Center for Semiconductor Physics in Nanostructures, The University of Oklahoma, Norman, Oklahoma 73069, USA}

\begin{abstract}
We have adapted R-matrix theory to calculate phonon scattering across systems of molecular to mesoscopic scale. 
The key novelty of this work is that the only required information about the scattering region are
its normal modes, which are evaluated only once for a system.
Thus, R-matrix theory is a computationally efficient and simple approach
to calculate phonon scattering in larger systems.
To validate and to demonstrate the applicability of the theory, we apply it to two systems:
a one dimensional chain of atoms and a graphene nanoribbon. In both cases, we discuss the effect of mass impurities on thermal transport.  
\end{abstract}

\pacs{63.22.-m; 65.80.-g; 65.80.Ck}
\maketitle

\section*{Nomenclature}
$a$-		lattice constant\\
$A$-		constant\\
$\mathcal{A}$-		matrix defined for convenience\\
$b$-		basis index\\
$\mathcal{B}$-		matrix defined for convenience\\
$c$-		phonon velocity\\
$h$-		Planck constant\\
$\hbar$-	reduced Planck constant ($\hbar=h/2\pi$)\\
IR-		interior region\\
$\mathcal{I}$-		identity matrix\\
$i$-		unitary imaginary number\\
$j$-		lattice index\\
$\mathcal{K}$-		mass normalized force constant matrix\\
$k_B$-		Boltzmann constant\\
$l$-		lead index(1,2)\\
$L$-		Lead\\
$\mathcal{L}_B$-		Bloch operator\\
$m$-		mass\\
$m_{Im}$-		Impurity mass\\
$\mathcal{M}$-		dynamical matrix\\
$n$-		integer use to number the eigen modes\\
$N_{uc}$-		number of atoms in a unit cell\\
$N_{\omega}$-		number of phonon branches at frequency $\omega$\\
$p$-		phonon branch\\
$q$-		wave vector\\
$Q$-		thermal current\\
$\mathcal{R}$-		R-matrix\\
$\mathcal{S}$-		scattering matrix\\
$t$-		time\\
$T$-		temperature\\
$u$-		displacement of an atom\\
$\uvec$-		displacement vector\\
$\vvec$-		eigen vector in the interior region\\
$V$-		many coordinate inter-atomic potential\\
$x$-		dimensionless variable\\ \\
\textbf{Greek Letters}\\
$\alpha$,$\beta$-		Cartesian degree of freedom\\
$\Gamma$-		transmission probability\\
$\epsvec$-			polarization vector of atoms in a unit cell in the lead	\\
$\eta$-		Bose-Einstein distribution function\\
$\omega$-		phonon frequency\\
$\tilde{\omega}$-	dimensionless phonon frequency\\
$\omega_D$-		cut off frequency\\
$\sigma$-		thermal conductance\\
$\tilde{\sigma}$-		dimensionless thermal conductance\\
$\lambda$-		eigen value in the interior region\\
$\tau$-		dimensionless temperature\\
$\phi$-			force constant\\
$\chivec$-		column vector defined for convenience\\
$\xivec$-		unit vector in $3N_{uc}$ dimensional space \\ \\
\textbf{Superscripts}\\
$\ast$ -		complex conjugate\\
$\dagger$ -		Conjugate transpose\\ \\

\section{Introduction \label{sec:Introduction}}

The rapid advancement of nanotechnology drives the fabrication of 
structures much smaller than the mean free path of  
electrons as well as phonons.
Electronic transport on the nanoscale  has been studied for over three decades and fascinating quantum effects have been observed \cite{Fowler,Peter,Wees}.
Phonon transport on this scale is of significant interest because of the increased
power dissipation  in nanoelectronics, which undermines the correct functionality of devices and limit their life time \cite{GChen,Balandin,Siemens,Adam,Arvind}. The need for better thermal conductivity has driven the search for a better understanding of thermal transport
on the nanometer level.

For example, the existence of a quantum of thermal conductance in mesoscopic
systems has been proposed \cite{rego} and demonstrated
experimentally \cite{seyler,schwab}. 
 In nanowires (width $<100$ nm),  phonon transport
is one dimensional at low temperatures due to the lateral confinement.\cite{arun}
In the ballistic limit, where the wire length is less than
or equal to phonon mean free path, thermal conductance reaches to its maximum $\pi^2 k_B^2 T/3 h$, which is called the quantum thermal conductance. In this limit, thermal
transport is largely determined by the boundaries (where the
boundaries are defined by the geometry of the device), interfaces and impurities.
Thus, it is essential to calculate the thermal conductivity including all the details of the
atomic constitutes to include the above effects.

Analogous nanoelectronic transport has been studied using the
Landauer and Buttiker formalism, in which the transport is determined
by the transmission probabilities \cite{lan,buti,Fowler,Peter,Wees}. The phonon
transport in these systems can also be treated by the very similar approach \cite{rego,blen}.  In this
context, the theories that was using initially in calculating
electronic transmission probabilities should be applicable to 
phonons.  For instance, the Green's function approach
has been successfully developed to calculate phonon transmission probabilities
\cite{Mingo,Zhang, NEGF1,NEGF2,NEGF3}.
 
R-matrix theory is a theoretical approach commonly used in nuclear and atomic physics to solve scattering problems \cite{wigner,AM1,AM2}.
The first development of the R-matrix theory draws back to the work of Wigner and Eisenbud in 
nuclear scattering in 1947.
Recently, this approach has been successfully developed to calculate electronic scattering
in mesoscopic quantum devices \cite{Lsm,uwulf,Thushari,varga}.
In this work, we present the first mathematical development of the R-matrix theory to calculate phonon transmission probabilities across a device of molecular to mesoscopic scale.

The key novelty of R-matrix theory for phonon is that the only required information about the scattering region 
is its normal modes, which is needed to evaluate only once for a system.
Normal modes are the collective oscillations of the lattice, which are orthonormal and 
form a complete basis in a finite region. 
These are calculated by diagonalising the dynamical matrix in the harmonic approximation of the interaction potential.
The harmonic approximation is well suited to study thermal transport in mesoscopic scale systems and low temperature transport, which is not accessible by the classical molecular dynamic simulations.
R-matrix theory is an efficient and simple way in calculating the phonon transmission through a nanoscale system.

 This paper is structured as follows: Section \ref{sec:Model}
 describes the mathematical formalism of the R-matrix theory.  An
 introduction to the model system  and the scattering picture of
 the lead is in section \ref{subsec:system}. In section
 \ref{sec.Rmat}, we construct the phonon version of the R-matrix.
 Then we solve for the scattering matrix ($S$) in section
 \ref{sec:ScatM}. To validate the theory and to demonstrate the applicability, we apply the theory for two type of systems
(Sec. \ref{application}). 
 First, we apply the theory to a  one dimensional
 chain of atoms. Then we calculate the  phonon transmission through a graphene nanoribbon. In both cases we discuss the effect of mass impurities to the thermal transport and a conclusion is given
in section \ref{conclude}.

\section{The R-matrix formalism
\label{sec:Model}}
\subsection{The system and the scattering picture 
\label{subsec:system}}
\begin{figure}[bt]
\centering
\includegraphics[scale=1.2]{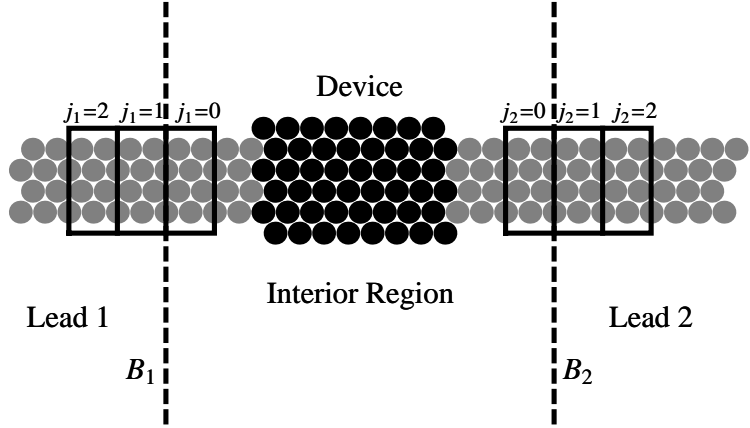}
\caption{{\protect\small Schematic of the model system having two semi-infinite leads connected to a device. The leads are assumed to be periodic and two boundaries, $B_1$ and $B_2$, are defined in the periodic region. The region between boundaries is  the 
{\it interior region} and the region outside is called the {\it asymptotic region}. The unit cells in the leads are depicted by the boxed regions. The lead lattice points are
indexed symmetrically by $j_l$, where $l=1(2)$ refers to the lead $1(2)$. }}
\label{uniform}
\end{figure}

The essence of 
R-matrix theory is to divide the system into 
an {\it interior region}(IR) and two or more {\it asymptotic regions}.
 The IR is the main scattering center of the system.
 The asymptotic  regions are the semi-infinite leads connected to the
  IR (Fig.\ref{uniform}).
  In the asymptotic regions, we assume the leads are narrow enough that the phonon transport in leads is strictly along the length direction of the leads.
 In the lateral direction there are confined phonon modes which is also called the phonon subbands.
 Thus, we define a extended unit cell, depicted by the boxed regions in the fig.\ref{uniform}, that has the periodicity in the length direction.
Although R-matrix theory can handle any number of leads, for simplicity we consider the case of only two leads, while it is straight forward to generalize to a system with many leads.
The IR and the leads are seperated by a set of boundaries in the periodic region, which we denote by $B_l$.
The unit cells in the  leads are indexed by $j_l$, where $l=1 (2)$ is the lead $1 (2)$.
 The lattice point $j_l=0$ is at the boundary $B_l$ in to the IR and $j_l$ increases positively outward the IR.
The heat current from lead $1$ to $2$ ($Q_{1,2}$) is given by the Landauer Formula (Eq.\ref{LanF}). \cite{rego,lan}
     \begin{equation}
    Q_{1,2} = \int_{\omega_{min}}^{\omega_{max}} \frac{d\omega}{2\pi}
    \hbar \omega \left(\eta_{L_1}(\omega) - \eta_{L_2} (\omega)
    \right) \Gamma_{1,2} (\omega), \label{LanF} \end{equation}
    where,  $ \eta_{L_l} (\omega) = \left\{ \exp \left( \hbar \omega
    / k_B T_{l} \right) -1\right\} ^{-1}$ is the Bose-Einstein
    distribution function of lead $l$,  $k_B$ is
the Boltzmann constant, $\hbar$ is the reduced Planck constant and $T_l$ is the temperature of lead $l$.
     The transmission probability of
    a phonon of angular frequency $ \omega$ from lead $1$ to $2$
    is given by $\Gamma_{1,2}(\omega)$.

 Since the leads are periodic there are well-defined phonons in each lead. The details of the phonons can be obtained solving
 the dynamical equation, \begin{equation}
  \omega ^2 \epsvec^p_{b,\alpha} = \sum _{b' \beta} \mathcal{M}(q)_{b
  b'}^{\alpha,\beta} \epsvec^p_{b',\beta}, \label{eq:dynamical}
  \end{equation}
over a unit cell, where the dynamical matrix $\mathcal{M}$ is,
\begin{equation}
  \mathcal{M}(q)_{b b'}^{\alpha,\beta} = \left( \sum _{j_l-j'_l}
  \phi _{j_lb,j'_lb'}^{\alpha,\beta} \frac{e^{i \qvec.(\Rvec_{j'_lb'}
  -\Rvec_{j_lb} )}}{\sqrt{m_b m_{b'}}}
  \right).  \label{DynMat} \end{equation}
 In the harmonic limit, force constants can be calculated by,
 \begin{equation}
  \phi _{j_lb,j'_lb'}^{\alpha,\beta}= \frac{\partial^2V}{\partial
  u_j^{\alpha} \partial u_{j'}^{\beta}},
\label{eq:calfc} \end{equation}
 which is evaluated  at the equilibrium positions of atoms.
The symbol $V$ refers to the many-coordinate inter-atomic potential.
 The coordinate of the atom $b$ in the unit cell $j_l$ is given by $\Rvec_{j_l,b}=\Rvec_{j_l}+\Rvec_b$, where $\Rvec_{j_l}$ is the position of the lattice point $j_l$ and $\Rvec_b$ is the position of the atom b within a unit cell. The Cartesian degree of freedom is represented by  $\alpha$ and $\beta$, $m_b$ is the mass of the atom $b$, $\qvec$ is the wave vector,  $u_j^{\alpha}$ is the displacement of the atom $j$ at the direction $\alpha$, and $\epsvec ^p_{b,\alpha}$ is the polarization amplitude of atom b in direction $\alpha$ in phonon branch $p$.
 There are $3N_{uc}$ phonon branches in each lead, where $N_{uc}$ is the number of atoms in the unit cell.
  
 We  can construct a scattering solution in the leads with these phonon modes. The scattering
solution is comprised of three components: the incoming wave, reflected wave and transmitted wave.
 We consider an incoming phonon of  frequency $\omega$, whose energy is $\hbar \omega$,  in lead $l_0$ and branch $p_0$. 
It can reflect in to any phonon branch ($p$) available at $\omega$ in lead $l_0$ and transmit in to any phonon branch ($p$) in lead $l$, which is different from $l_0$. 
 We assume all the scattering processes are elastic and the anharmonic interactions are neglected.
 As we defined our leads symmerically,  the incoming waves are going at the direction of decreasing  the lead lattice index $j_l$ and the scattered waves (both reflected and transmitted) are going at the increasing direction of the lattice index.
 Thus, the incident and scattered components of the scattering solution can be written in terms of plane waves as, $\chivec^p(-\qvec^p(\omega))$ and $\chivec^{p}(\qvec^p(\omega))$ respectively, where $\chivec^p(\qvec^p(\omega))$ is a vector  whose entries are the values of $\epsvec^p_{b} \exp(-i\omega t + i\qvec^p(\omega).\Rvec_{j_l,b})/\sqrt{m_b}$ of all the atoms in a unit cell.
 Now the scattering solution ($\uvec^l_{l_0,p_0}$) of the lead $l$ for above described scattering event can be written as,
 \begin{equation}
 \uvec^l_{l_0,p_0}(j_l)= \sum_p \left( \chivec^p(-\qvec^p(\omega)) \delta_{p,p_0} \delta_{l,l_0} + \chivec^p(\qvec^p(\omega)) \mathcal{S}^{l,p}_{l_0,p_0}(\omega) \right),
 \label{scatsol}
 \end{equation}
 where,the  first part inside the summation represents an incoming wave in branch $p_0$ in lead $l_0$ and the second part is the scattered wave.
 The scattering amplitude in branch $p$ lead $l$ is given by $\mathcal{S}^{l,p}_{l_0,p_0}(\omega)$, which is a complex number.
 The scattering amplitudes form a $2N_{\omega}\times 2N_{\omega}$  scattering matrix $\mathcal{S}(\omega)$, where the $N_{\omega}$ is the number of phonon branches at frequency ($\omega$). 
The matrix elements of $\mathcal{S}$, when $l=l_0$ are the reflection coefficients and $l\neq l_0$ are the transmission coefficients. 

 From the scattering amplitudes, we can calculate the transmission probabilities.
 The transmission probability is the ratio of the energy fluxes carried by each vibrational wave.
 The transmission probability from branch $p_0$ in lead $l_0$ to branch $p$ in lead $l (\neq l_0)$ can be expressed as,  $ \Gamma^{l p}_{l_0 p_0}(\omega) = \mathcal{S}^{l,p \ast}_{l_0,p_0}(\omega) \mathcal{S}^{l,p}_{l_0,p_0}(\omega) \it{c}^p(\omega)/\it{c}^{p_0}(\omega)$, where 
 $\it{c}^p(\omega)$ is the phonon velocity of the branch $p$ at frequency $\omega$ \cite{blen}.  
Thus the transmission probability from lead $l_0$ to $l(\neq l_0) $ is,
\begin{equation}
\Gamma_{l l_0}(\omega) = \sum_{p p_0} \Gamma ^{l p}_{l_0 p_0}(\omega).
\label{Tprob}
\end{equation}

 \subsection{R-matrix \label{sec.Rmat}}
 
 The dynamics of the entire system can be described by an infinite set of coupled equations, which are usually written as matrix form,
 \begin{equation}
 \left(\omega^2 \mathcal{I} + \mathcal{K} \right) \uvec=0,
 \label{eq:bigsys}
 \end{equation}
 where $\mathcal{I}$ is the identity matrix, $\uvec$ is the displacement vector and $\mathcal{K}$ is the mass normalized force constant matrix.
 The matrix elements of $\mathcal{K}$  can be calculated  as,
 \begin{equation}
 \mathcal{K}_{j,j'}^{\alpha,\beta}=\frac{1}{\sqrt{m_jm_{j'}}} \phi^{\alpha,\beta}_{j j'},
 \label{eq.kmat}
 \end{equation}
 where $j$ refer to a common index of atoms used to label all the atoms in the system.
 The matrix $\mathcal{K}$ can be expressed more descriptively using the block matrices: $\mathcal{K}_{IR}$ is defined only in the finite {\it IR}, $ \mathcal{K}_{L_l} $  is defined in the lead $l$ that is infinite, and $K_{IR,L_l}$ is the coupling between the interior region and the leads.
 \begin{equation}
\mathcal{K} \uvec =\left(\begin{array}{ccc}  \mathcal{K}_{L_1} & \mathcal{K}_{L_1,IR} & 0 \\
 \mathcal{K}_{IR,L_1} & \mathcal{K}_{IR} &  \mathcal{K}_{IR,L_2} \\  0 & \mathcal{K}_{L_2,IR} & \mathcal{K}_{L_2} \end{array} \right) \left(\begin{array}{c} \uvec_{L_1} \\ \uvec_{IR} \\ \uvec_{L_2}\end{array} \right), 
 \label{eq:blockm}
 \end{equation}
where $\uvec_{L_l}$ and $\uvec_{IR}$ are the displacement vectors of atoms in  lead $l$ and the {\it IR}.

It is natural to  describe the IR by 
normal modes of vibrations.
However, in our scattering problem the IR  is connected
to semi-infinite leads and thus we can not solve for the normal modes.
The solution is to solve the equations of motion for a different, but
related physical system.
When solving the continuum Schrodinger equation 
in a finite region,  a ``Bloch operator" is included to 
solve the problem of non-Hermiticity of the kinetic energy operator \cite{Thushari,varga}.
The resulting Hamiltonian is called the 
``Bloch Hamiltonian", which is explicitly Hermitian inside a finite region.
In the phonon problem we define a ``Bloch operator ($\mathcal{L}_B$)", 
which includes the couplings between the IR and the leads as follows,
\begin{equation}
\mathcal{L_B}=\left( \begin{array}{ccc} 0&\mathcal{K}_{L_1,IR}&0\\ \mathcal{K}_{IR,L_1} & 0 & \mathcal{K}_{IR,L_2} \\ 0 &  \mathcal{K}_{L_2,IR} & 0 \end{array} \right).
\label{eq:LB}
\end{equation}
The term  $\mathcal{L_B} \uvec$ gives the coupling forces between the IR and the leads.
It can involve interactions of any finite range, e.g. nearest neighbor, next nearest neighbor, etc.  
We subtract the term $\mathcal{L_B} \uvec$ from the both side of the eq.\ref{eq:bigsys} yielding,
\begin{equation}
\left(\omega^2 I + \mathcal{K} -\mathcal{L_B}\right) \uvec=-\mathcal{L_B} \uvec.
\label{eq:decoupled}
\end{equation}
The matrix ($\mathcal{K}-\mathcal{L}_B$) on the left is called the 
``Bloch dynamical matrix'', in which the coupling between the two regions are removed.
The eq.\ref{eq:decoupled} can be written in block matrices as follows,
\begin{equation}
\omega^2 \left(\begin{array}{c} \uvec_{L_1} \\ \uvec_{IR} \\ \uvec_{L_2}\end{array} \right) + \left(\begin{array}{ccc}  \mathcal{K}_{L_1} & 0 & 0 \\
 0 & \mathcal{K}_{IR} &  0 \\  0 & 0 & \mathcal{K}_{L_2} \end{array} \right) \left(\begin{array}{c} \uvec_{L_1} \\ \uvec_{IR} \\ \uvec_{L_2}\end{array} \right) = -\left( \begin{array}{ccc} 0&\mathcal{K}_{L_1,IR}&0\\ \mathcal{K}_{IR,L_1} & 0 & \mathcal{K}_{IR,L_2} \\ 0 &  \mathcal{K}_{L_2,IR} & 0 \end{array} \right) \left(\begin{array}{c} \uvec_{L_1} \\ \uvec_{IR} \\ \uvec_{L_2}\end{array} \right).
 \label{blocheq}
\end{equation}
We can extract the center row of the eq.\ref{blocheq}, which describe the IR,
\begin{equation}
\left( \omega^2 \uvec_{IR} + \mathcal{K}_{IR} \uvec_{IR}\right) = -\sum_l \mathcal{L_B}^l \uvec_{L_l},
\label{eq:dcseperate}
\end{equation}
where $\mathcal{L_B}^l=\mathcal{K}_{IR,L_l}$.
In the ``Bloch dynamical matrix" IR can be solved independently. We find eigenvalues ($\lambda_n$) and eigenvectors ($\vvec^{(n)}$)
 of $\mathcal{K}_{IR}$ according to,
\begin{equation}
\left(\lambda_n^2 + K_{IR}\right) \vvec^{(n)} =0.
\label{eq:IRonly}
\end{equation}
These eigenvectors are orthonormal and form a complete set as the mass normalized force constant matrix is always Hermitian.
A general solution in the IR can be expanded  using the above eigen vectors as follows, 
   \begin{equation}
   \uvec_{IR} = \sum_n A_n  \vvec^{(n)}.
    \label{IRgs}    
    \end{equation}
    By plugging this in to the left hand side of the equation (\ref{eq:dcseperate}) yields,
      \begin{equation}
     \sum_n\left(\lambda_n^2- \omega^2\right) A_n  \vvec^{(n)}=\sum_l \mathcal{L_B}^l \uvec_{L_l}.
    \label{IRgs2}    
    \end{equation}
   By taking the inner product with $\mathcal{V}^{(n) \dagger}$ ($\dagger$ refers to the conjugate transpose) we find,    
    \begin{equation}
      A_n= \sum_l \frac{ \vvec^{(n) \dagger} \mathcal{L_B}^l \uvec_{L_l}}{\lambda_n^2-\omega^2}.
    \label{IRgsAn}    
    \end{equation}
   Finally, the scattering solution in the IR can be expressed as,
    \begin{equation}
      \uvec_{IR} = \sum_{n,l}  \vvec^{(n)} \frac{ \vvec^{(n) \dagger} \left( \mathcal{L_B}^l \uvec_{L_l} \right)}{\lambda_n^2-\omega^2}.
    \label{IRgsF}    
    \end{equation}

    According to the eq.\ref{eq:blockm} and \ref{eq:LB}, Bloch operator of lead $l$, $\mathcal{L_B}^l$, has infinite number of columns and $3N$ rows, where $N$ is the number of atoms in the IR.
  However there are non-zero couplings only between a subset of atoms  in the IR  with a subset of atoms in the leads,
  in the visinity of the boundary $B_l$.
  Since the boundary is defined in the periodic region, we assume that the minimum size of the coupling regions to be unit cells at either side of the boundary. 
This is a reasonable (but not essential) 
 assumption for most practical situations.
 Therefore we can construct a matrix of dimensions $3N_{uc} \times 3N_{uc}$  that contains the non-zero couplings in  $\mathcal{L_B}^l$. 
This effective part of the Bloch operator is denoted by $\bar{\mathcal{L_B}}^l$. 
Now we replace $\uvec_{L_l}$ by $\uvec^l_{l_0 p_0}(j_l=1)$, of which the entries are the displacements of the atoms in the unit cell at lattice point $j_l =1$. Similarly $\left(\vvec^{(n)}\right)_l$  contain the normal modes amplitudes of the   atoms in the unit cell at $j_l=0$.
Now we can replace  $ \vvec^{n \dagger} \left( \mathcal{L_B}^l \uvec_{L_l} \right)$ by $\left( \vvec^{(n) \dagger}\right)_l \left( \bar{\mathcal{L_B}}^l \uvec_{l_0 p_0}^l(j_l=1) \right)$ in eq.\ref{IRgsF}. 

In the electron transport problem, the R-matrix relates the value of the  wave function to its normal derivative at the boundary of the scattering region 
and the normal derivative at the boundary serves as the Bloch operator. \cite{Thushari}
The phonon version of the R-matrix ($\mathcal{R}$) can be defined as follows,
\begin{equation}
\uvec^{l''}(j_{l''}=0)= \sum_{s'', s',l'} \xivec^{s''} \mathcal{R}_{l'',l'}^{s'',s'} \xivec^{s' \dagger} \bar{\mathcal{L_B}}^{l'} \uvec^{l'}(j_{l'}=1),
\label{Reqn}
\end{equation}
\begin{equation}
\mathcal{R}_{l'',l'}^{s'',s'}= \sum_{n} \frac{\xivec^{s'' \dagger} (\vvec^{(n)})_{l''} (\vvec^{(n)})_{l'}^{\dagger} \xivec^{s'}}{\lambda_n^2-\omega^2},
\label{Rmat}
\end{equation}
where $\xivec^s$'s are a set of unit vectors that are orthonormal and complete in the $3 N_{uc}$ dimensional space.
These vectors  can be chosen arbitrarily and a possible choice is given in appendix \ref{unitvec}.

The eq.\ref{Reqn} is obtained by matching the lead solution and the interior region solution at the unit cell at $j_l=0$.
In general, the matching is done on the 
surface atoms that share the both regions.
The choice of the unit cell is adapted for the convenience of explanation.
This particular choice is valid for some practical cases such as zigzag graphene nanoribbons.
Moreover, the incorporation of unit vectors $\xivec^s$ is not essential. 
It allows to calculate the matrix elements of the $\mathcal{R}$ before hand so that it can be used in the scattering matrix calculation.
Alternatively, one can keep the summation in eq.\ref{Reqn} only over on ``$n$" and ``$l'$" without including the unit vectors $\xivec^s$ and proceed.

 \subsection{Scattering Matrix
 \label{sec:ScatM}}
 The scattering matrix ($\mathcal{S}$) can be obtained by plugging the explicit form of the scattering solution in the lead ($\uvec^l_{l_0 p_0}$) from eq.\ref{scatsol} in to the R-matrix equation (\ref{Reqn}). We can express the scattering matrix as,
\begin{equation}
\mathcal{S}(\omega)=-\left[\mathcal{A}(\qvec)-\mathcal{B}(\qvec)\right]^{-1}.\left[\mathcal{A}(-\qvec)-\mathcal{B}(-\qvec) \right],
\label{Smat}
\end{equation}
 where the matrix elements of $\mathcal{A}$ and $\mathcal{B}$ are given by,
\begin{equation}
\begin{array}{c}
\mathcal{A}^{l''p''}_{l'p'} =  \epsvec^{p'' \dagger} \chivec^{p'}(\qvec^{p'};\Rvec_{j_l=0}) \delta_{l''l'}, \\
\mathcal{B}_{l' p'}^{l'' p''}=\sum _{s'' s'}   \epsvec^{p'' \dagger} \xivec^{s''}\mathcal{R}^{s'' s'}_{l'' l'}\left(\xivec^{s' \dagger} \bar{\mathcal{L_B}}^{l'} \chivec^{p'}(\qvec^{p'};\Rvec_{j_l=1})\right).
\end{array}
\label{Tmat}
\end{equation}
Equation (\ref{Smat}) is our key expression for the scattering matrix.
 A derivation of this is given in appendix \ref{solveS}. 
 The matrices $\mathcal{A}$ and $\mathcal{B}$ are independent of the number of atoms in the  IR.
 It is the R-matrix that represent the IR in the scattering calculation.
 In order to construct the  matrices $\mathcal{A}$, $\mathcal{B}$  and $\mathcal{R}$, we only need to calculate  for the normal modes of the  IR in the Bloch dynamical matrix and the lead phonons.
 The appealing feature is that these details are needed to be calculated only once for a system.

\section{Applications \label{application}}

\subsection{One dimensional chain of atoms \label{sec:1dresults}}
First, we calculated the transmission probabilities for a one dimensional germanium channel connected to silicon leads using R-matrix theory and could reproduce the results in reference \cite{NEGF2} which was calculated using the nonequilibrium Green's function method.
Here, we use R-matrix theory to calculate the phonon transmission of  a  one-dimensional(1D)  chain of atoms that has one atom in the unit cell and 
investigate the effect of mass impurities to the thermal transport.
In particular, we study three cases: a single mass impurity, diatomic chain, and a random mass impurity distribution connected to two semi infinite uniform leads.

The atomic masses are $m_j=1.0$ for all $j$ and the couplig constants are given by,
\begin{equation}
\phi_{ij}=\left \{ \begin{array}{ccc} 2 & \mbox{for} & i=j \\
-1 & \mbox{for} & j=i\pm 1\\
0 & \;  & otherwise \end{array} \right.
\end{equation}
 The IR of the system consists of 25 atoms.
The Bloch operator
($\mathcal{L}_B^l$) for this 1D system is a single number, which is the force constant
between the atom in the boundary and the  adjacent atom in the lead.
 We consider only the
longitudinal  vibrational modes.  

\begin{figure}[bt] \centering \begin{tabular}{c}\includegraphics[scale=0.80]{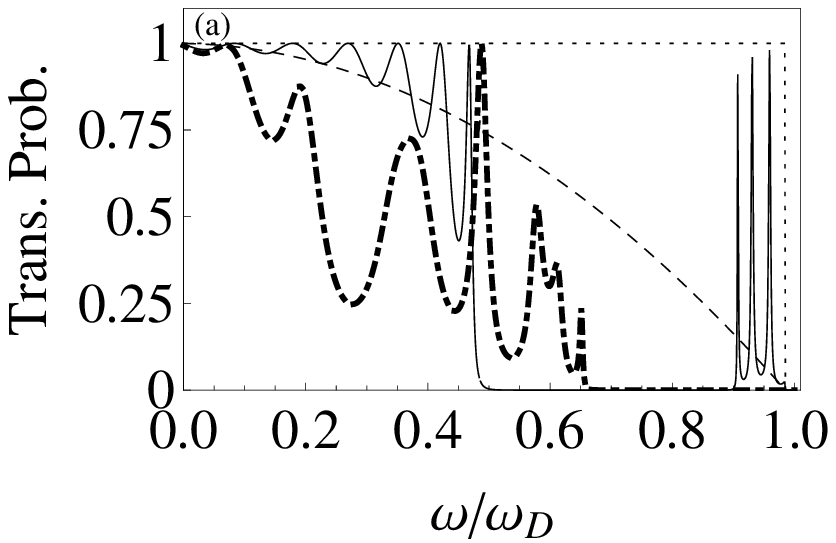} \\
\includegraphics[scale=0.80]{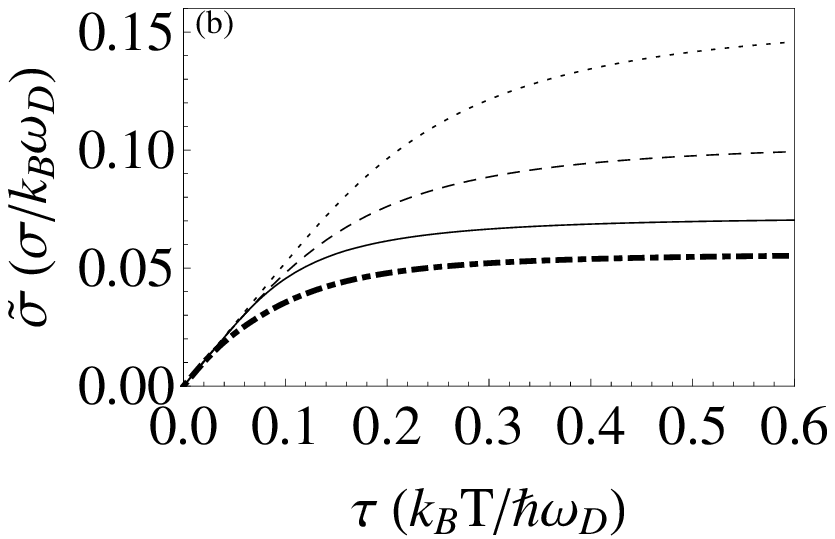}\end{tabular}
\caption{{\protect\small (a)Transmission probability ($\Gamma$)
as a function of dimensionless frequency ($\omega /\omega_D$). (b) Dimensionless thermal conductance ($\tilde{\sigma}$) as a function of dimensionless temperature ($\tau$). The
dotted line is for the uniform chain. Dashed line is for the inclusion
of single impurity of  $m_{Im}^\one=2.0$ and periodic diatomic structure
of 7 unit cells with masses $m_{Im}^\one$ and $m_{Im}^\two=0.6$ is given
by the solid line. The thick dotted dashed line is for the random
distribution of 14 atoms with masses $m_{Im}^\one$ and $m_{Im}^\two$. }}
\label{Imp-1} \end{figure}

The dotted line in fig.\ref{Imp-1}a shows the transmission probability
for uniform chain. It gives unit transmission for all the frequencies
up to the cut off $\omega_D$ .
 Mass impurity is made by changing the mass of an atom in the middle
 of the interior region to $m_{Im}^\one=2.0$.
  The dashed line in fig.\ref{Imp-1}a shows the transmission
  probability calculated in the presence of a single mass impurity.
  In the semi-classical approach impurity point defect scattering is modeled by a  
  Rayleigh type expression \cite{Ray1,Ray2} that is proportional to $\omega^4$. 
  We observe a similar trend in the transmission probability for scattering from 
  a single mass impurity in 1D.
 Then we incorporate another mass of  $m_{Im}^\two=0.6$ with the
 $m_{Im}^\one$ forming a periodic diatomic structure in the middle of
 the interior region. The number of unit cells included is seven.
 The corresponding transmission is plotted by the solid line in
 fig.\ref{Imp-1}a. We can see the opening of a forbidden region
 having zero transmission. Next the above two masses are randomly
 distributed  replacing 14 atoms in the interior region. In this
 case the transmission is depicted by the dotted dashed line.

Now we calculate the thermal conductance as a function of temperature.
 In the linear response regime, where  $\Delta T \ll T$,  $\Delta T= |T_1-T_2|$ and $T=\left(T_1 + T_2 \right)/2$,   the thermal conductance ($\sigma$) of the system can be expressed as follows,   
  \begin{equation}
  \sigma = \frac{k_B^2 T}{h} \int_{x_{min}}^{x_{max}} dx \frac{\Gamma(x) x^2 e^x}{\left(e^x -1 \right)^2},
   \label{LanTC}    
   \end{equation} 
   where, $x=\frac{\hbar \omega}{k_B T}$.
 The dimensionless thermal conductance ($\tilde{\sigma}=\sigma /(k_B \omega_D)$) is defined as,
\begin{equation}
   \tilde{\sigma} = \frac{1}{2 \pi} \int_{0}^{1} d\tilde{\omega} \frac{\Gamma(\tilde{\omega}) \tilde{\omega}^2 e^{\tilde{\omega}/\tau}}{\tau^2\left(e^{\tilde{\omega}/\tau} -1 \right)^2},
    \label{DlessL}    
    \end{equation}
where $\tau=\left(k_B T/\hbar \omega_D\right)$ ,  $\omega_D$ is the cut off frequency, which is equal to 2.0 in this problem , $\tilde{\omega}=\omega/\omega_D$.

The  dimensionless thermal conductance for previously described  cases is plotted in fig.\ref{Imp-1}b. 
The random distribution of masses shows the lowest thermal conductance.
The random distributions could creates localized states suppressing the phonon transport whereas the periodic structures are promising for travelling waves.
 
 \subsection{Graphene nanoribbons \label{GNR}}
Graphene nanoribbons (GNRs) are $10\sim100$ nm wide strips of graphene with high aspect ratios.
Graphene is a promising material for nanoscale applications due to its exceptional electronic  \cite{elecp} and thermal properties \cite{therp,susp,therp2}.
Thermal conductivity of graphene is found to be greater than $3000$ W/mK in recent experiments \cite{therp,susp,therp2}.
Moreover it has been recently revealed, both theoretically \cite{flex} and experimentally \cite{supot}, that the flexural (out of plane) vibrational modes are the governing energy carriers in graphene.  We apply R-matrix theory to calculate the transmission of flexural vibrational modes of a GNR.

 \begin{figure}[bt]
\centering
\includegraphics[scale=0.70]{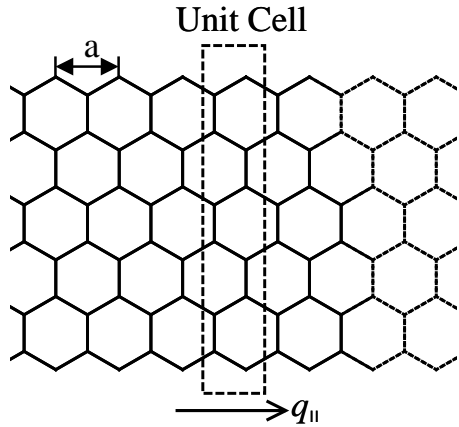}
\caption{{\protect\small Schematic of the a zigzag graphene nanoribbon (ZGNR) of six zigzag chains in width.  A unit cell is depicted by the boxed region. The lattice constant in the leads is denoted by ``$a$''. The horizontal arrow shows the direction of $q_{\shortparallel}$.}}
\label{fig.GNR}
\end{figure}
Figure \ref{fig.GNR} shows a schematic of a zigzag graphene nanoribbon (ZGNR) of six zigzag chains in width.
   A unit cell is depicted by the boxed region and the lattice constant is given by $a$.   
    The force constants are calculated using a second generation Brenner potential\cite{brenner}.    
    When considering interactions up to the next nearest neighbors, there are non-zero couplings only between the atoms in adjacent unit cells. This support the assumption we make in defining the effective part of the Bloch operator in sec.\ref{sec.Rmat}. First we construct the $\mathcal{K}$ matrix as in eq.\ref{eq.kmat} including few unit cells in the leads. Then we extract the $\mathcal{K}_{IR}$ and $\mathcal{L}_B^l$ as describe in the eq.\ref{eq:blockm} and \ref{eq:LB}.
    
\begin{figure}[bt]
\centering
\includegraphics[scale=0.80]{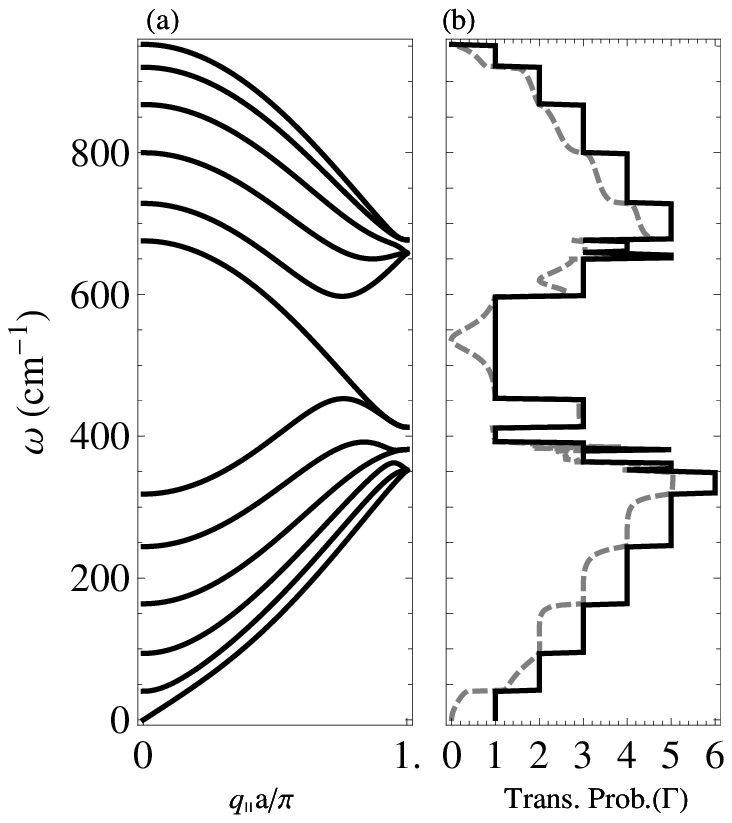} 
\caption{{\protect\small Phonon dispersion (a) of flexural vibrational modes of  the ZGNR. The transmission probability ($\Gamma$) calculated for the perfect ZGNR from R-matrix theory (b). The $\Gamma$ count the number of phonon branches available at each frequency. The dashed line refers to the case where a larger mass in the middle of the IR. The frequency is measured  in the units of cm$^{-1}$. }}
\label{fig.Tperfect}
\end{figure}
 \begin{figure}
\centering
\includegraphics[scale=0.700]{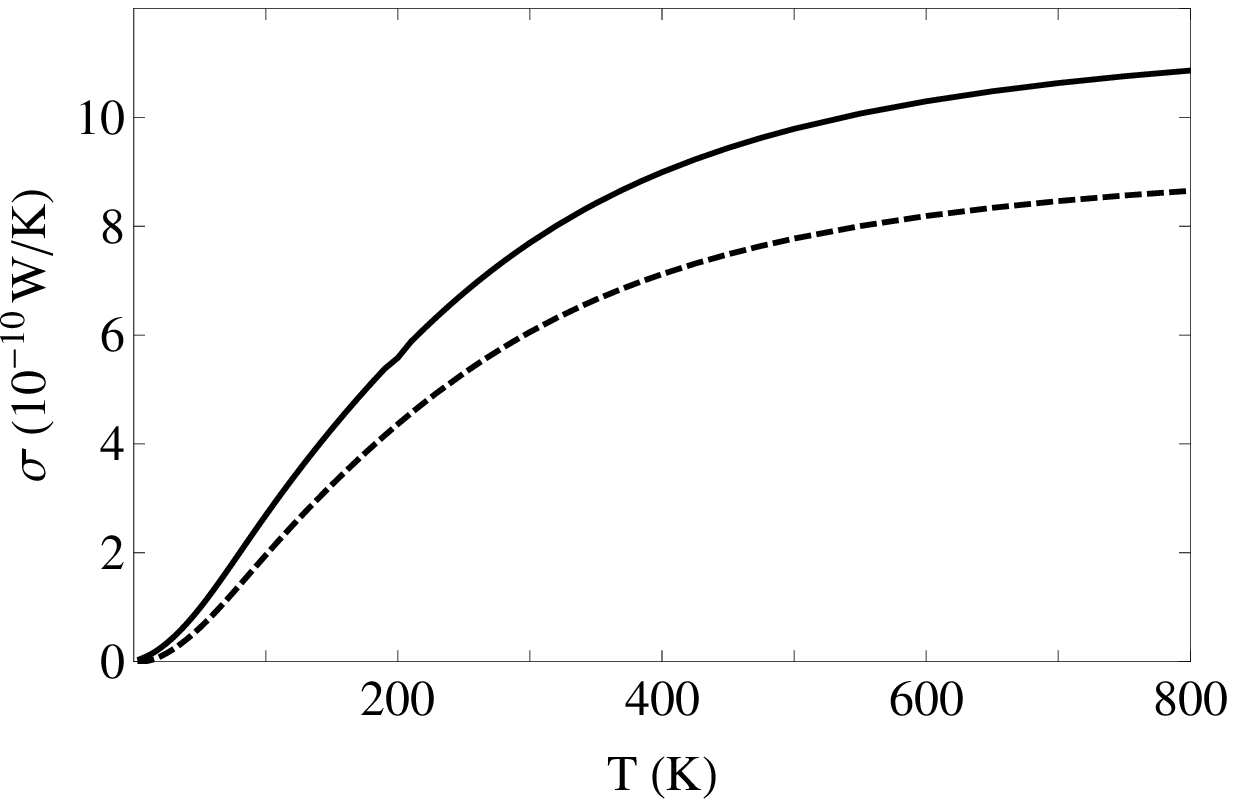}
\caption{{\protect\small  Thermal conductance due to out of plane vibrations of the ZGNR as a function of temperature. Solid line is for the perfect structure and the dashed line is the result due to the larger mass in the center.}}
\label{fig.TMImp}
\end{figure}
\begin{figure}[bt]
\centering
 \includegraphics[scale=0.700]{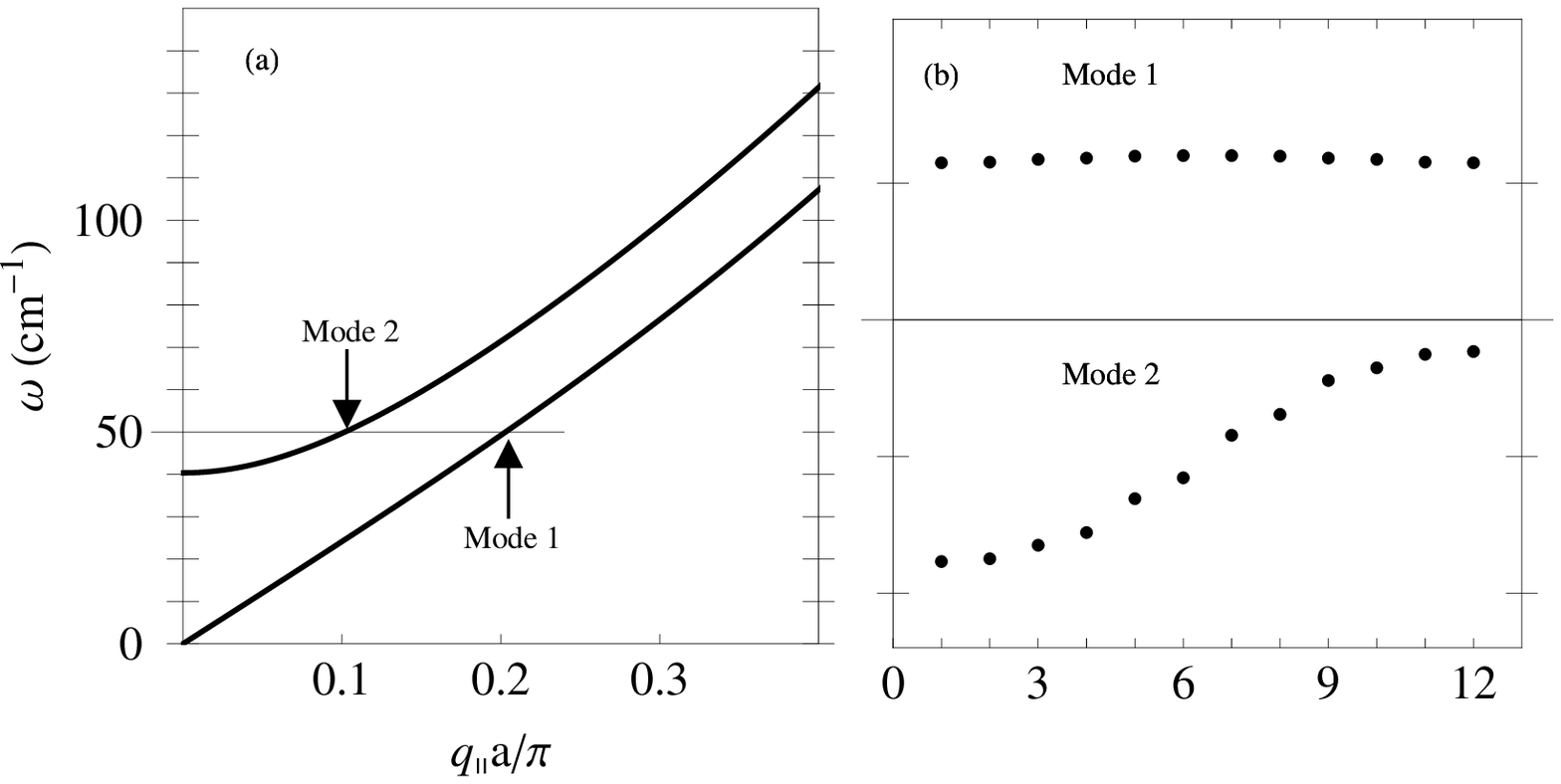} 
\caption{{\protect\small Two lowest lying flexural phonon branches of the ZGNR. The fig.(a) shows the phonon dispersion in the low frequency region. Figure (b) shows the polarization profile of atoms in an unit cell calculated at $50$ cm$^{-1}$ of the acoustic branch (mode 1) and the  twisting branch (mode 2).}}
\label{fig.2LLM}
\end{figure}

First, we calculate the phonon dispersion of flexural modes according to the eq.\ref{eq:dynamical}. Figure \ref{fig.Tperfect}a shows the phonon dispersion along the direction of the ribbon length, where $q_{\shortparallel}$ is the wave vector projected on to the length direction.
There are $12$ phonon branches available due to the fact that the unit cell has 12 atoms. 
These phonon branches represent  different laterally confined phonon subbands that are specially available in nanowires.
The acoustic branch is also available in large graphene sheets. 
The lowest lying optical branch, which has a twisting character (mode 2 in fig.\ref{fig.2LLM}),  can only  exist in nanoribbons.
 In larger graphene sheets twisting modes do not present since it is
energetically costly.
 Figure \ref{fig.Tperfect}b shows the calculated transmission probability ($\Gamma(\omega)$)  from the R-matrix theory.
  The $\Gamma$ counts the  number of phonon branches available at each frequency.
  Similar behavior of the transmission probability has been reported by the
non-equilibrium Green's function calculations.
  
  We incorporate a larger mass (mass impurity of $m_{Im}=100 m_{C^{12}}$) replacing one carbon atom in the center of the interior region.  We keep about 20 unit cells (240 atoms) in the interior region.  Although this is not a realistic situation, could be analogous to a strong coupling to a heavy molecule.
  The calculated transmission probability for this case is shown by the dashed line in fig.\ref{fig.Tperfect}b.
  We observe that the overall transmission of phonons  is reduced due to the heavy mass.
  In fig.\ref{fig.TMImp} we plot the thermal conductance ($\sigma$) as a function of the temperature.
  Thermal conductance of the ZGNR with the mass impurity (dashed line) is considerably lower than that of uniform ZGNR (solid line).
 \begin{figure}
\centering
\begin{tabular}{c}
\includegraphics[scale=0.70]{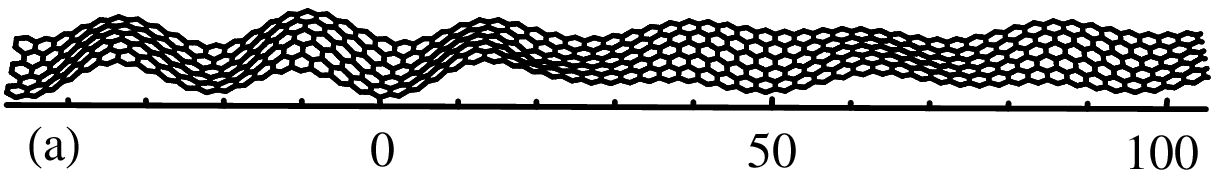}\\
\includegraphics[scale=0.70]{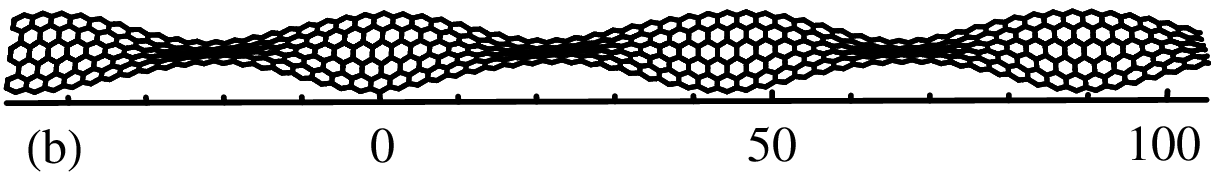} \end{tabular}
\caption{{\protect\small Displacement wave in three regions (lead 1, interior region and lead 2) is sketched together when a larger mass is present in the center of the interior region. The left boundary of the interior region ($B_1$) starts from the origin of the longitudinal coordinate (marked in units of \AA) and the right boundary ($B_2$) is at 20 unit cells away from the $B_1$ (at $40$ \AA). The heavy mass is at the $10^{th}$ unit cell from the $B_1$ (at $20$ \AA). The sketch at top (a) is for waves coming in mode 1 at $\omega=50$ cm$^{-1}$. The sketch in bottom (b) is for the waves coming in mode 2 at the same frequency. }}
\label{fig.Displot}
\end{figure}

In R-matrix theory we are working on the real space displacement waves. This allows us to easily sketch the profile of the displacement waves during the scattering. This visualization is important to gain clear insight on to the scattering process and tuning the mode specific transmission of phonons.
 In fig.\ref{fig.Displot}, we sketch the displacement wave in the three regions: lead 1, interior region and lead 2, at frequency $50$ cm$^{-1}$ when the mass impurity is in the center of the interior region.
  We use the eq.\ref{scatsol} for the leads and the eq.\ref{IRgsF} for the interior region. 
  The displacement waves in three regions matches nicely at the boundaries.
  There are two phonon subbands available at this frequency as shown in fig.\ref{fig.2LLM}.
  The sketch in fig.\ref{fig.Displot}a is for the waves coming in mode 1, which is the acoustic mode.
  We observe that the propagation of the acoustic mode is considerably attenuated with the presence of the mass impurity. 
  In the acoustic mode, all the atoms in the unit cell vibrate in-phase. Thus, the  mass impurity prevent this in-phase motion.
  However, we observe that the twisting mode is not affected by the presence of the mass impurity in the middle as implied by the fig.\ref{fig.Displot}b. This could be attributed to the asymmetric polarization profile of the twisting mode (mode 2 in fig.\ref{fig.2LLM}).

\section{Conclusion \label{conclude}}

 We have adapted the R-matrix
  theory  to calculate phonon scattering  on the atomic level.
  The major advantage of the R-matrix approach is the reduced
  computational time. 
  Moreover the ability to express the scattering matrix by a simple
  analytical expression is also an advantage of this approach as is
the capability of easily plotting the scattering displacement wave.
 The application to devices with
three or more leads is straightforward and only involves increasing  the 
range of the lead index.
We discuss results applying R-matrix theory 
 to a 1D atomic chain and a graphene nanoribbon. 
 We observe a lower thermal conductivity for a random distribution of masses than the periodic arrangement in a 1D atomic chain.
 For a single mass impurity, the profile of the transmission probability shows a trend similar to the Rayleigh type scattering.
 However,  phonon transmission through GNRs in the presence of a mass impurity shows strong dependence on the mode profile of the laterally  confined vibrational modes. 
 
 Furthermore, there are a number of important differences between R-matrix theory
for electrons and for phonons.  First, because the electron case is defined
over a continuous spatial degree of freedom, there are an infinite number
of eigenstates of the Bloch Hamiltonian. This means that any solution 
involves a high energy cut-off given by the largest eigenvalue considered
in the Bloch Hamiltonian.
In the phonon case, there is no 
sense to oscillations that have a wavelength shorter than the distance
between atoms in a single unit cell.  Therefore, there are no more than
$3N$ modes associated with the interior region, and such convergence
is not an issue. The electron case also suffers convergence issues related
to the choice of basis used for solutions of the Bloch Hamiltonian.  
Since the scattering solution can have any amplitude and derivative on 
the boundary, it is important that the basis functions share the same
feature. This motivates a variety of different approaches to choosing a
basis, such as variational basis functions. In contrast, in the phonon 
case we merely solve for the normal modes of the system, and there is no
such ambiguity.

The electron problem has the advantage of well-defined boundaries and
a clear potential energy.  Setting up the Bloch dynamical matrix requires
knowledge of the force constants (eq.(\ref{eq:calfc})) evaluated at
the equilibrium position of the atom.  This may require simulating the
system and allowing it to relax, minimizing the internal strain.
Similarly, to define the ``lead solutions'' we must assume that the lead
lattices are semi-infinite and uniform.  In real systems the interior
region may exert a strain that distorts the leads;  the interior region
must be large enough so that plane wave solutions in the leads are valid.

The electronic scattering problem involves many ``channels."  In atomic
physics these are usually different angular momentum states for the
scattering electron, while in the mesoscopic device case they correspond to
different subbands in a lead.  Such inter-subband scattering is present
in phonons as well.  Electronic {\it interband}  transitions  require
a periodic lattice
potential in the leads and/or the interior region, so that it is meaningful
to discuss electron bands.
The phonon system may also have  
different bands if the unit cell has a multi-atom basis.   This opens up the
possibility of phonon transduction, where optical phonons can be turned
into
acoustic phonons and vice versa, through the suitable design of an interior
region.  

\textbf{acknowledgements}\\
This project was supported in part by the US National Science Foundation\\
under Grant~\mbox{MRSEC DMR-0080054} and AFOSR Grant~\mbox{FA9550-10-10031 EPSCoR 2009}.

-----------------------------------------------------------------------------

\begin{appendix}
\small
\section{A choice of  vectors $\xivec^s$ \label{unitvec}}
The set of vectors $\xivec^s$ can be chosen arbitrarily. One possible choice  is,
\begin{equation}
\begin{array}{cccc}
\xivec^{(1)}=\left( \begin{array}{c} 1 \\ 0 \\0\\0 \\0\\ \vdots  \end{array} \right), & 
\xivec^{(2)}=\left( \begin{array}{c} 0 \\ 1 \\0\\0 \\ 0\\ \vdots  \end{array} \right), &
\cdots ,&
\xivec^{(n)}=\left( \begin{array}{c} 0 \\ \vdots \\0\\1 \\0\\ \vdots  \end{array} \right).
\end{array}
\end{equation}
\vspace*{5pt}

\section{Solving for scattering matrix ($\mathcal{S}$) \label{solveS}}
In this appendix, we discuss the algebra of solving  for the scattering matrix ($\mathcal{S}$).
By plugging $\uvec^{l''}_{l_0 p_0}(j_l=0)$ and $\uvec^{l''}_{l_0 p_0}(j_l=1)$ from eq.\ref{scatsol} to eq.\ref{Reqn} and taking the projection of both side on to $\epsvec^{p'' }$, we can obtain,
\begin{equation}
\begin{array}{c}
\sum_{p'''} \epsvec^{p''\dagger }  \chivec^{p'''}(-\qvec^{p'''};\Rvec_{j_l=0}) \delta_{p''' p_0}
\delta_{l'' l_0} + \epsvec^{p'' }  \chivec^{p'''}(\qvec^{p'''};\Rvec_{j_l=0})
\mathcal{S}^{l'' p'''}_{l_0 p_0} = \\ \sum_{s'' s' l' p'} \epsvec^{p''\dagger} \xivec^{s''}\mathcal{R}^{s'' s'}_{l'' l'} \left[
\left(\xivec^{s' \dagger} \mathcal{L}_B^{l'}  \chivec^{p'}(-\qvec^{p'};\Rvec_{j_l=1})\right) \delta_{p' p_0} \delta_{l' l_0} + \left(\xivec^{s' \dagger} \mathcal{L}_B^{l'}  \chivec^{p'}(\qvec^{p'};\Rvec_{j_l=1}) \right) \mathcal{S}^{l' p'}_{l_0 p_0} \right] \end{array}
\label{eq.Ap1}
\end{equation}
We define a matrix ($\mathcal{B}$) for the convinent as follows,
\begin{equation}
\mathcal{B}_{l' p'}^{l'' p''}=\sum _{s'' s'}   \epsvec^{p''\dagger} \xivec^{s''}\mathcal{R}^{s'' s'}_{l'' l'}\left(\xivec^{s' \dagger} \mathcal{L}_B^{l'} \chivec^{p'}(\qvec^{p'};\Rvec_{j_l=1})\right) 
\label{Tmat}
\end{equation}
We rewrite eq.\ref{eq.Ap1} rearranging the parts and inserting $\mathcal{B}$.
\begin{equation}
\sum_{p'''}\epsvec^{p''\dagger} \chivec^{p'''}(-\qvec^{p'''};\Rvec_{j_l=0}) \delta_{p''' p_0} \delta_{l'' l_0} - \sum_{l' p'} \mathcal{B}_{l' p'}^{l'' p''}(-\qvec^{p'}) \delta_{p' p_0} \delta_{l' l_0} = \sum_{l' p'} \mathcal{B}_{l' p'}^{l'' p''}(\qvec^{p'}) \mathcal{S}^{l' p'}_{l_0 p_0} - \sum_{p'''} \epsvec^{p'''\dagger } \chivec^{p''}(\Rvec_{j_l=0}) \mathcal{S}^{l'' p''}_{l_0 p_0}.
\label{eq.A3}
\end{equation}
Let's focus on the R.H.S. of the eq.\ref{eq.A3}. It can be written as,
\begin{equation}
\begin{array}{c}
\sum_{l' p'}  \left( \mathcal{B}_{ l' p'}^{l'' p''}(\qvec^{p'}) \mathcal{S}^{l'
p'}_{l_0 p_0} -  \delta_{l''l'} \epsvec^{p''\dagger} \chivec^{p'}(-\qvec^{p'};\Rvec_{j_l=0}) \mathcal{S}^{l' p'}_{l_0 p_0} \right) \\
= -\left(\mathcal{A}(\qvec) -\mathcal{B}(\qvec)\right) \mathcal{S},
\end{array}
\end{equation}
where $\mathcal{A}^{l''p''}_{l'p'} =  \epsvec^{p''\dagger} \chivec^{p'}(\qvec^{p'};\Rvec_{j_l=0}) \delta_{l''l'}$.
Similarly, we can show the L.H.S. of  eq.\ref{eq.A3} is $\left( \mathcal{A}(-\qvec) - \mathcal{B}(-\qvec)\right)$.
Thus, we can finally write,
\begin{equation}
\mathcal{S}=-\left[\mathcal{A}(\qvec)-\mathcal{B}(\qvec)\right]^{-1}.\left[\mathcal{A}(-\qvec)-\mathcal{B}(-\qvec) \right]
\end{equation}

\end{appendix}

\begin{thebibliography}{99}

\bibitem{Fowler}  A. B. Fowler, A. Hartstein,  and R. A. Webb, Conductance in Restricted-Dimensionality Accumulation Layers, {\it Phys. Rev. Lett.}, Vol. 48, p. 196, 1982.
\bibitem{Peter} F. M. Peeters, Quantum Hall Resistance in the Quasi-One-Dimensional Electron Gas
, {\it Phys. Rev. Lett.}, Vol. 61, p. 589, 1988.
\bibitem{Wees} B.  J. van Wees, H. van Houten, C. W. J. Beenakker,  J. G. Williamson, L. P. Kouwenhoven, D. van der Marel, 
and C. T. Foxon, Quantized conductance of point contacts in a two-dimensional electron gas
, {\it Phys. Rev. Lett.}, Vol. 60, p. 848, 1988. 

\bibitem{GChen} G. Chen, Phonon heat conduction in nanostructures,{\it Int. J. Therm. Sci.} Vol. 39, p. 471, 2000.
\bibitem{Balandin} J. Zou, and  A. Balandin, Phonon heat conduction in a semiconductor nanowire, {\it J. Appl. Phys.}, Vol. 89, p. 2932, 2000.
\bibitem{Siemens} M. E. Siemens, Q. Li,  R. Yang, K. A. Nelson, E. H. Anderson, M. M. Murnane, and H. C. Kapteyn,Quasi-ballistic thermal transport from nanoscale interfaces observed using ultrafast coherent soft X-ray beams, {\it Nature Materials}, Vol.9, p. 26, 2009.
\bibitem{Adam} A. Christensen and S. Graham, Multiscale lattice Boltzmann modeling of phonon
transport in crystalline semiconductor materials, {\it Numer. Heat Transfer, Part B}, Vol. 57, p. 89, 2010.
\bibitem{Arvind} A. Pattamatta and C. K. Madnia, A comparative study of two-temperature and Boltzmann transport models for electron-phonon nonequilibrium, {\it Numer. Heat Transfer, Part A}, Vol. 55, p. 611, 2009.

\bibitem{rego} L. G. C. Rego and G. Kirczenow, Quantized Thermal Conductance of Dielectric Quantum Wires, {\it Phys. Rev. Lett.} Vol. 81, p. 232, 1998.
\bibitem{seyler} J. Seyler and M.N. Wybourne, Acoustic waveguide modes observed in electrically heated metal wires, {\it Phys. Rev. Lett.} Vol. 69, p. 1427,1992. 
\bibitem{schwab} K.Schwab, E.A.Henriksen, J.M.Worlock, and M.L.Roukes, Measurement of the quantum of thermal conductance, {\it Nature}, Vol. 404, p. 974, 2000.
\bibitem{arun} D. Li, Y. Wu, P. Kim, L. Shi, P. Yang, and  A. Majumdar,Thermal conductivity of individual silicon nanowires, {\it Appl. Phys. Lett.}, Vol. 83, p.2934, 2003.

\bibitem{lan} R. Landauer, Spatial Variation of Currents and Fields Due to Localized Scatterers in Metallic Conduction, {\it IBM J. Res. Dev.}, Vol. 1, p. 223, 1957.
\bibitem{buti} M. Buttiker, Four-Terminal Phase-Coherent Conductance, {\it Phys. Rev. Lett.}, Vol. 57, p. 1761, 1986.


\bibitem{blen} M. P. Blencowe, Quantum energy flow in mesoscopic dielectric structures
, {\it Phys. Rev. B}, Vol. 59, p. 4992, 1999.

\bibitem{Mingo} N. Mingo, and L. Yang,
Phonon transport in nanowires coated with an amorphous material: An atomistic Green's function approach, {\it Phys. Rev. B.}, Vol. 68, p. 245406, 2003.
\bibitem{Zhang} W. Zhang, T. S. Fisher, and N. Mingo, The Atomistic Green's Function Method: An Efficient Simulation Approach for Nanoscale Phonon Transport, {\it Numer. Heat Transfer, Part B}, Vol. 51, p. 333, 2007.
\bibitem{NEGF1}  J. -S. Wang, J. Wang, and N. Zeng,
Nonequilibrium Green's function approach to mesoscopic thermal transport, {\it Phys. Rev. B}, Vol. 74, p. 33408, 2006.
\bibitem{NEGF2} P. Hopkins, P. Norris, M. Tsegaye, and A. Ghosh,Extracting phonon thermal conductance
across atomic junctions: Nonequilibrium Green's function approach compared to semiclassical methods,
{\it Jour. App. Phys.}, Vol. 106, p. 63503, 2009.
\bibitem{NEGF3} Y. chen, T. Jayasekera, A. Calzolari, K. W. Kim, and M. B. Nardelli, Thermoelectric properties of graphene nanoribbons, junctions and superlattices, {\it J. Phys.:Condens. Matter}, Vol. 22, p. 372202, 2010.

\bibitem{wigner} E. P. Wigner, and L. Eisenbud, Higher Angular Momenta and Long Range Interaction in Resonance Reactions, {\it Phys. Rev.}, Vol. 72, p. 29, 1947.
\bibitem{AM1} P. G. Burke, A. Hibbert, and W. D. Robb, Electron scattering by complex atoms 
, {\it J. Phys. B: At. Mol. Phys.}, Vol. 4, p. 153, 1971.
\bibitem{AM2} B. I. Schneider, and P. J. Hay, Elastic scattering of electrons from F2: An R-matrix calculation, {\it Phys. Rev. A}, Vol. 13, p. 2049, 1976.
\bibitem{Lsm} L. Smrcka, R-matrix and the coherent transport in mesoscopic systems, {\it Superlattices Microstruct.}, Vol. 8, p. 221, 1990.
\bibitem{uwulf} U. Wulf, J. Kucera, P. N. Racec, and E. Sigmund, Transport through quantum systems in the R-matrix formalism
, {\it Phys. Rev. B}, Vol. 58, p. 16209, 1998.
\bibitem{Thushari} T. Jayasekera,  K. Mullen, and M. A. Morrison, R-matrix theory for magnetotransport properties in semiconductor devices
, {\it Phys. Rev. B}, Vol. 74, p. 235308, 2006.
\bibitem{varga} K. Varga,R-matrix calculation of Bloch states for scattering and transport problems
, {\it Phys. Rev. B}, Vol. 80 , p. 085102, 2009.


\bibitem{Ray1} J. W. Vandersande, and C. Wood, The thermal conductivity of insulators and semiconductors, {\it Contemp. Phys.}, Vol. 27, p. 117, 1986.
\bibitem{Ray2} S. Barman, and G. P. Srivastava, Thermal conductivity of suspended GaAs nanostructures: Theoretical study, {\it Phys. Rev. B}, Vol. 73, p. 205308, 2006.


\bibitem{elecp} K. S. Novoselov, A. K. Geim, S. V. Morozov, D. Jiang, M. I. Katsnelson, I. V. Grigorieva, S. V. Dubonos, and  A. A. Firsov, Two-dimensional gas of massless Dirac fermions in graphene, {\it Nature }, Vol. 438, p. 197, 2005.


\bibitem{therp} D. L. Nika, S. Ghosh, E. P. Pokatilov, and A. A. Balandin, Lattice thermal conductivity of graphene flakes: Comparison with bulk graphite, {\it Appl. Phys. Lett.}, Vol. 94, p. 203103, 2009.
\bibitem{susp} A. A. Balandin, S. Ghosh, W. Bao, I. Calizo,
D. Teweldebrhan, F. Miao, and C. N. Lau, Superior Thermal Conductivity of Single-Layer Graphene
, {\it Nano. Lett.}, Vol. 8, p. 902, 2008.
\bibitem{therp2} W. Cai, A. L. Moore, Y. Zhu, X. Li, S. Chen, L. Shi, and R. S. Ruoff, Thermal Transport in Suspended and Supported Monolayer Graphene Grown by Chemical Vapor Deposition, 
{\it Nano. Lett.}, Vol. 10, p. 1645, 2010.

\bibitem{flex} L. Lindsay, D. A. Broido, and N. Mingo, Flexural phonons and thermal transport in graphene, {\it  Phys. Rev. B}, Vol. 82, p.115427, 2010.
\bibitem{supot} J. H. Seol, I. Jo, A. L. Moore, L. Lindsay, Z. H. Aitken, M. T. Pettes, X. Li, Z. Yao, R. Huang, D. Broido, et al., Two-Dimensional Phonon Transport in Supported Graphene,
 {\it Science}, Vol. 328, p. 213, 2010.


\bibitem{brenner} D. W. Brenner, Empirical potential for hydrocarbons for use in simulating the chemical vapor deposition of diamond films, {\it Phys.Rev. B}, Vol. 42, p. 9458, 1990.


\end{thebibliography}

\clearpage


\end{document}